\documentclass[12pt,thmsa]{article}
\usepackage{sw20aip}

\input{tcilatex}
\input tcilatex
\begin{document}

\title{Recent verifications of a new relativity principle and a new gravitational
theory based on properties of light}
\author{Rafael A. Vera\thanks{%
rvera@udec.cl} \\
Departamento de F\'{i}sica. \\
Facultad de Ciencias F\'{i}sicas y Matem\'{a}ticas.\\
Universidad de Concepci\'{o}n. Concepcion.\\
Chile.}
\date{}
\maketitle

\begin{abstract}
Recent astronomical observations verify the new scenario resulting from new
conservation laws and a new relativity principle fixed either by dual
properties of light or by new gravitational (G) tests and the Einstein's
equivalence principle. This scenario is radically different from the
classical one. During a free fall, the relative masses of free bodies, with
respect to the observer at rest in a G field, remain constants. The energy
released during the stops in different G potentials (GP) comes not from the
G field but from the bodies. The relative properties of the bodies at rest
with respect to the observer depend on the differences of GP between the
bodies and the observer. The increase of GP due to universe expansion
expands bodies in identical proportion. The universe age may be infinite.
Its entropy is conserved because the new black hole, without singularity,
after absorbing radiation, explodes regenerating new gas that transforms
dark galaxies into luminous ones. Galaxies evolve, indefinitely in closed
cycles with luminous and dark periods. All of their phases are found
anywhere in the universe. They solve fundamental dilemmas like dark matter
and radiation backgrounds
\end{abstract}

\section{INTRODUCTION}

In the Einstein's Centennial Symposium on Fundamental Physics (Bogot\'{a},
1979), the dual properties of light have been used as a crucial test for
gravitational (G) theories [1]. Thus it is proved, according to wave
continuity, that the relative frequency of an electromagnetic wave-train
traveling in a static G field, with respect to any well-defined observer at
rest the field, must be conserved\footnote{%
From wave continuity, during the trip, \textit{the number N of waves between
any two waves of a wave train is conserved}. On the other hand, \textit{each
wave takes the same time interval, of a clock of constant frequency, to
travel between any two points A and B of a static G field}. Then the
relative frequency of the waves crossing the point B with respect to the
clock of the observer at A is identical to the frequency of the waves
crossing the point A, with respect to the same clock}. This new
``conservation law'' is in clear contradiction with the Einstein's G field
energy hypothesis (GFEH) according to which the photons would change their
frequencies, during their trips, due to a hypothetical energy exchange with
the G field.

This new law has been generalized for bodies on the base of ``\textit{a new
theory based on a particle model made up of photons in stationary states''}
[2]. This model is justified from the fact that, ``according to the
Einstein's equivalence principle (EEP), all of the well-defined parts of a
system that must obey the same inertial and gravitational laws''. Since the
minimum well-defined part of a system is a photon in stationary state
between other well-defined parts of the same system, then such photon must
have the same inertial and gravitational properties of the other parts of
such system. Then in principle \textit{the inertial and gravitational
properties of bodies can be derived, independently on the traditional
hypotheses, by using a particle model made up of a photon in stationary state%
}.

Effectively, so far, the relative properties of such model, derived
theoretically, do correspond with special relativity (SR), quantum mechanics
and with all of the classical tests for G theories [2], [3]. They can be
summarized into a new ``\textit{relative mass-energy conservation law}'' and
``\textit{a non-equivalence principle}'' (NEP)\textit{\ for bodies in
different GP}. According to them, the relative mass-energy and the relative
frequencies of a body (or a photon) traveling freely in a G field, with
respect to any observer at rest in a constant GP, remain constants. If the
body stops in different G potentials it releases, locally, different
energies, and, therefore, its final relative properties are linearly
dependent on its corresponding difference of GP with respect to the
observer. \textit{The proportional differences of frequencies, mass-energies
and lengths of the NL body, compared with the local ones, at the observer's
potential, are just equal to their respective differences of GP}.

The NEP has been successfully tested with all of the classical tests for G
theories [1] [2]. Below, it has been directly verified, independently on the
particle model, from a new test based on the condition of consistency of the
EEP with the results of genuine GTD experiments that don't depend on the
frequency of any photons traveling between the clocks. Further verifications
come from the agreement of the new astrophysical context predicted from such
principle and some recent astronomical observations.

\section{A NEW TEST FOR THE NEW PRINCIPLE}

From the results of the genuine GTD experiments it is concluded that ``%
\textit{the proportional differences of frequency of the standard clocks at
rest in different GP are just equal to the differences of GP between them}%
''. Then, to describe the real physical differences of a non-local (NL)
clock at some NL position B with respect to some observer in a position A, a
new formalism must be used in which the positions A and B must be clearly
stated. Here, and in the works cited above, the observer's position, which
is most important, has been indicated by a subscript. Thus the results of
GTD experiments, reduced to static conditions, can be stated in the form:

\begin{equation}  \label{1}
\frac{\nu _{A}(0,B)-\nu _{A}(0,A)}{\nu _{A}(0,A)}=\Delta \phi _{A}(B)=\frac{
\Delta E_{A}(B)}{m_{A}(0,A)}
\end{equation}

The first member is called ``\textit{the proportional difference of
frequency of the NL clock at B with respect to the (local) clock at A}''.
This one turns out to be just equal to the difference of GP between B and A,
called $\Delta \phi _{A}(B)$, which is defined in (1) as the proportional
energy released by the clock, or any other small test body, after a free
fall from B and a stop at A, compared with its local rest mass-energy at A.
Notice that the masses and the energies are expressed in energy units.

From (1), the clock at B is physically different compared with the clock at
A, i.e., the CH is wrong.

Then the NEP can be directly derived from the general condition of
correspondence of (1) with the EEP. For such purpose, let $\nu _{A}(0,A)$, $%
m_{A}(0,A)$ and $\lambda _{A}(0,A)$ be the generic symbols for the ``\textit{%
\ local}'' values of any frequency, mass-energy, length or wavelength,
respectively, of any isolated part or photon in stationary state of a system
at rest at A, with respect to the observer at rest at A. From the EEP, such
values are related to each other by constants that don't depend on the
differences GP of the local system with respect to other systems.

\begin{equation}  \label{2}
\nu _{A}(0,A):m_{A}(0,A):\lambda _{A}(0,A)=C1:C2:C3
\end{equation}

Then (1) can be consistent with (2) only if:

\begin{equation}  \label{3}
\frac{\Delta \nu _{A}(0,B)}{\nu _{A}(0,A)}=\frac{\Delta \nu _{A}(0,B)}{\nu
_{A}(0,A)}=\frac{\Delta \nu _{A}(0,B)}{\nu _{A}(0,A)}=\Delta \phi _{A}(B)
\end{equation}

In which $\Delta \nu _{A}(0,B)$ $=\nu _{A}(0,B)-\nu _{A}(0,A)$

From (3), the proportional differences of the basic properties of standard
bodies at rest at B with respect to the observer at A, are just equal to the
difference of GP between B and A.

This is a direct verification of the NEP originally derived, step by step,
from properties of light[1],[2].

Notice that the NEP is more general than the EEP because it is also valid
for NL cases in which bodies and observers are in different GP. Obviously,
it corresponds with the EEP in any local case in which the difference of GP
between the NL object and the observer tends to zero.

To the contrary of the EEP, the NEP shows that some real (absolute) physical
differences exist between standard bodies in different GP, i.e., that the CH
is wrong. Then \textit{the current relationships between quantities measured
by observers at rest in different GP are inhomogeneous because their
reference standards are physically different compared to each other\footnote{%
This turns out to be the source of most of the problems in gravitation,
astrophysics and cosmology, as shown below}}.

From the 2nd and the last member of (3):

\begin{equation}  \label{4}
\Delta E_{A}(B)=\Delta m_{A}(0,B)=m_{A}(0,B)-m_{A}(0,A)
\end{equation}

\textit{The energy released during G work comes not from the G field but
from the transformation of a fraction of the mass of a body into free energy}%
.

For example, during a free fall from B to A, special relativity can be
applied locally at A, just before the stop at A. If $m_{A}(V,A)$ is the
relativistic mass of the moving body with respect to the observer at A, then:

\begin{equation}  \label{5}
\Delta E_{A}(B)=m_{A}(V,B)-m_{A}(0,A)
\end{equation}

From (4) and (5),

\begin{equation}  \label{6}
m_{A}(V,A)-m_{A}(0,B)
\end{equation}

\textit{During a free fall, the relativistic mass of a NL body, with respect
to an observer at rest in a well-defined potential, remains constant}. This
is just the relative mass-energy conservation law for NL bodies with respect
to inertial observers at rest in some well-defined GP. Then, to the contrary
of GR, the G field does not exchange energy with bodies and photons.

\subsection{Gravity is a Refraction Phenomenon}

If $\nu _{A}(0,B)$ is the relative frequency of a standing electromagnetic
wave of a system at rest at B, with respect to the observer at A, and if $%
\lambda _{A}(0,B)$ is its relative wavelength, then the relative speed of NL
light at B with respect to the observer at A is:

\begin{equation}  \label{7}
c_{A}(B)=\nu _{A}(0,B)\lambda _{A}(0,B)
\end{equation}

Then, from (3) and (7),

\begin{equation}  \label{8}
\frac{\Delta \nu _{A}(0,B)}{\nu _{A}(0,A)}=\frac{\Delta \lambda _{A}(0,B)}{%
\lambda _{A}(0,A)}=\frac{\Delta c_{A}(B)}{2c_{A}(B)}=\Delta \phi _{A}(B)
\end{equation}

From (8) it is clear that the G field is a gradient of the relative
refraction index of the NL space with respect to observers in a fixed GP,
i.e., \textit{that gravitation is a refraction phenomenon due to a gradient
of the relative speed of light}. This is obvious, anyway, in the \textit{G
lens effect}.

Notice that this is a self-consistency test for the new theory because in
any refraction phenomenon there is a momentum exchange without energy
exchange between the photons and the dielectric. Thus the refraction
occurring during the round trips of the model's standing waves accounts for
both the momentum exchange and the lack of energy exchange between the
bodies and radiations and the G field.

\subsection{Discussion}

According to the CH, the relative mass of a body at rest with respect to the
observer would be independent on the difference of GP between the body and
the observer. Thus the CH rules out the possibility that bodies can give up
the energy released after a change of GP. Then \textit{a reason for which
general relativity (GR) accounts for the classical tests for G theories is
that the proportional errors of the CH are compensated by those of the GFEH
introduced by Einstein}. But this happens only when the two hypotheses are
used. The real errors of each hypothesis do stand out when just one of them
is used.

For example, GR accounts for the results of experiments on G red shift (GRS)
in which it is tacitly assumed that the relative emission frequency of the
standard atoms located in different GP are the same with respect to each
other, which is the CH. At the same time it is assumed that, during the trip
of the photons, their frequencies would change due to some hypothetical
energy exchange between the photons and the G field, which is the GFEH. Thus
the errors of the two hypotheses, of the same magnitude and opposite signs,
are exactly compensated.

On the other hand, from just wave continuity, the GFEH is necessarily wrong.
The same holds for the CH because, from just GTD experiments, the relative
frequencies of the standard clocks in different GP are different with
respect to each other. Such differences existed \textit{before} any photon
can travel between the clocks.

The referee of an important journal of physics has argued that: in a GTD
experiment, the two clocks located in different GP should run with the same
frequency because a photon traveling between the clocks would be red shifted
during its trip due to the work done by the G field. Obviously, he did nor
realize in that the results of such experiments don't depend on the
frequency of any photon traveling between the clocks. Anyway, such argument
is equivalent to say that the CH should be true just because the GFEH should
be true, i.e., that the first hypothesis of GR should be true just because
the second hypothesis of GR should be true. This is a clear vicious circle.
The point here is that each of these two hypotheses is wrong.

\section{THE NEW UNIVERSE FIXED BY THE NEP}

In the standard cosmology it is assumed that the measurement rods would not
expand after the increase of GP coming from universe expansion. Such
hypothesis is in contradiction with the phenomenon of ``gravitational
expansion'' fixed by (8).

It may be argued that the strong forces within the structure of particles
would prevent such expansion. This is not true because (8) comes from the
EEP and experiments whose results are independent on the existence of such
forces.

Assume, as a ``\textit{trial hypothesis''}, that bodies don't expand during
universe expansion. Thus, after a time interval $\Delta t$, the proportional
increase of every relative distance in the universe, with respect to the
observer, would be the same and equal to $H\Delta t$. Then it is simple to
find that the proportional increase of GP at the observer's place would be,

\begin{equation}  \label{9}
\Delta \phi =\frac{\Delta r}{r}=H\Delta t
\end{equation}

From (3), (9), the proportional expansion of a body is:

\begin{equation}  \label{10}
\frac{\Delta \lambda }{\lambda }=\Delta \phi =\frac{\Delta r}{r}=H\Delta t
\end{equation}

Thus the ``trial hypothesis'' is wrong because it is not possible to find a
real reference standard, or reference frame, that does not expand in just
the same proportion as any other distance of the universe. Thus,
paradoxically, a global universe expansion cannot increase any measured
distance, velocity or cosmological red shift because matter must expand in
just the same proportion. From the relative (measurable) viewpoint, the
average universe must look like it was static, for ever, i.e., its age can
be infinite.

On the other hand, from the ``relative mass-energy conservation law'', the
relative mass-energies would also be conserved during a universe expansion.
Anyway, this is obvious from the lack of real energy exchange between the
space and the free bodies.

The same conclusion comes out, even more obviously, after emulating every
part of the universe by particle models made up of photons in stationary
states. Then it is found, from the Huygens's principle, that the photons and
particles would be the result of constructive interferences of wavelets that
are traveling rather indefinitely in the universe. Thus a universe expansion
would stretch every wavelet in just the same proportion, which would not
change any ratio between its parts\footnote{%
Notice that the ''wavelets'' turn out to be most important for a simplified
and unified understanding a wide range of phenomena occurring in the
universe.}.

\subsection{The New Kind of Black Hole}

The strict linearity of the new equations eliminates the odd singularity
produced by the non linear equations of GR. Thus the new kind of ``linear
black hole'' (LBH) is just a huge ``macronucleus'' that \textit{obeys
ordinary nuclear laws }[1], [2], [3]. From (8), due to the very low relative
speed of light, the relative mass of its nucleons must be very low compared
with the ones far away from the LBH.

The high gradient of the relative refraction index of the space around the
LBH would prevent the escape of individual photons and nucleons, after the
phenomenon of critical reflection. On the other hand such gradient would
increase its cross section for capturing radiation coming from all over the
universe.

Then the average mass-energy per neutron in the LBH, with respect to an
observer at infinite, must increase with the time until it becomes equal to
the mass of a free neutron far away from it. This is equivalent to hot
plasma confined in a small volume, by its own G field. In such critical
state the LBH can decay, rather adiabatically, thus generating a new
primeval gas of composition similar to that predicted for the hot \textit{%
Big-Bang} theory. This gas can start a new evolution cycle of matter within
a small volume of the universe, like a galaxy.

\subsection{Evolution Cycles of the Galaxies}

From the new properties of the black holes and from the longer age of the
universe it is inferred that, in one way or another, an appreciable fraction
of the matter in a galaxy must be evolving in rather closed cycles between
the states of gas and LBH and vice versa.

In the long run, a galaxy, after radiation emission, must end as a ``\textit{%
dark galaxy}'' made up of one or a couple of LBH surrounded by a set of dead
stars, planets and planetesimals, in a space rather free of plasmas. Such
galaxy cannot collapse because their uncharged bodies cannot emit gravitons%
\footnote{%
This is not the case of binary pulsars in which there is plasma falling in
combination of magnetic and G fields that transforms G energy into electric
energy. These charged bodies do radiate ''photons'' that would take away
angular momentum thus making believe in that they are ''gravitons''. So far
the experiments for detecting gravitons have failed.}. After an out gassing
period, and after a long period of radiation absorption, its LBH should
explode, in chain, thus generating new primeval gas that can transform the
dark galaxy into a luminous one.

Thus a galaxy should evolve in rather closed cycles, with luminous and dark
periods, rather indefinitely. Its luminous periods must end when the whole
galaxy has run out of available energies. The dark periods would end when
the LBH has absorbed energy enough to explode

\subsection{The New Model of Star Formation}

In the new scenario, the primordial gas comes not from a universal Big-Bang
but from LBH explosions occurring mostly in the central regions of dark
galaxies. Most of this gas would be absorbed by the cool bodies of the dark
galaxy. The evolution states of the new luminous bodies must also depend on
the evolution states of the dark bodies from which they come from.

Then this new scenario accounts for the existence, within the same galaxy,
of two main kinds of stars in different evolution states: the really ``new''
stars coming gas captured by low mass bodies such as planets and
planetesimals of the dark period, and the ``renewed'' stars coming from gas
captured by dead stars or neutron stars of the same dark period. Thus,
strictly, the last ones are older than the first ones. The new stars should
have low densities and low temperatures, and the last ones would have higher
densities and higher temperatures. Paradoxically, the properties of the
``new'' stars correspond with the so called ``old'' stars. Those of the
``renewed'' ones correspond with the so called ``young'' stars.

Notice that a renewed star can also be formed from gas compressed by the
strong G field gradient of an earlier ``neutron star''. Since the G binding
energy of the neutrons can be of a higher order of magnitude than the
nuclear one, and then in principle such star should show higher temperatures
(bluer) than the original stars where it comes from. Thus a neutron star is
likely to be growing up rather hidden in the center of some bluer stars of
the new galaxy.

In principle, a naked neutron star can react with the ions falling on it,
mainly along magnetic lines, according to ``neutron stripping reactions''.
Due to the higher G binding energy of the neutrons, the protons must be
rejected along its magnetic axis. They would normally be stopped by the
external gas shell. For thin shells, they can pass through them, which would
account for the ``cosmic jets'' that in turn would become sources of radio
waves. They account for the low proportion of neutrons in the ``cosmic
rays''.

Then the direct fall of matter on naked neutron stars should also produce
high energy radiation like cosmic rays, variable X-rays and gamma bursts.
The continuous fall of plasma on a super massive neutron star, in a very low
GP, would produce a star-like object of high G red shift that can be
misinterpreted as a cosmological red shift. They would correspond with the
so called \textit{quasi stellar radio sources, i.e., quasars.}

\subsection{The New Model of Galaxy Evolution}

A luminous period of a galaxy should start after a chain of LBH explosions
occurring in a dark galaxy. The dark bodies, after absorbing new gas, would
form two main sets of stars: the strictly ``new'' stars, most of them with
random angular momentum orientations, and a small number of ``renewed''
stars with higher proportions of preferred angular momentum orientations
that come from the dark period. Thus the new galaxy, of a rather spherical
shape, would correspond with ``elliptical galaxies'' that have maximum
proportions of new gas and minimum proportion of metal contamination and
dark bodies.

During its luminous period the renewed stars can capture gas clouds and
steal gas from near stars of lower densities. In the long run, they may
become more powerful stars while the other less massive stars may become
dark ones. Thus, in the average, the renewed of stars may stay luminous
longer than the lower density stars. They must show up the higher proportion
of the preferred orientation of such kind of stars. Thus the luminous volume
of the elliptical galaxy should take the form of a disk, or a spiral,
surrounded by a rather spherical halo of dark bodies.

In the meanwhile, the more dense and massive stars should drift towards the
center of the galaxy, in counter current with lower mass stars. They must
grow in mass by capturing gas and other lower mass bodies. Thus, a cluster
of massive neutron stars should grow up in the central region of the galaxy.
They should become sources of high energy radiations coming from
nuclear-gravitational reactions. Thus the spiral galaxy should turn into an
AGN.

In the long run, the central cluster of neutron stars should be reduced to a
single super massive LBH, or a binary one. This one would be a star-like
object that would be still capturing recycled gas of the galaxy. It would be
surrounded by a shell of hot plasma supported its magnetic field. Due to its
low GP, with respect to the observer, it should emit light with a high G red
shift. The fall of positive ions along its polar regions would produce
narrow jets of high energy protons (\textit{cosmic rays}) along the magnetic
axis, according to nuclear stripping reactions, which would become source of 
\textit{radio waves}. This star-like object is consistent with the genuine
quasi stellar radio sources (quasars) of high red shift.

During the dark period of a galaxy, after an out gassing period, the
uncharged bodies cannot loose the presumed gravitons because they don't
exchange energy with G fields. This property would prevent the collapse of
the dark galaxy. Thus, after a long period of radiation absorption, the
nucleons of its most massive LBH can reach the critical relative mass after
which it may explode thus starting a new galaxy cycle.

Since the universe age would be rather infinite, then, statistically, ``%
\textit{the different evolution phases of the galaxies and clusters should
be rather uniformly distributed in the universe in the proportions fixed by
the periods of their corresponding phases}''.

Due to the small average energy absorption rate of a dark galaxy, compared
with the high emission rate in its luminous period, the dark period
necessary for recovering the energy lost during the luminous period must be
of a higher order of magnitude than the luminous period. Then,
statistically, the average proportion of dark galaxies, compared with the
luminous ones, must also be a high number, i.e., ``most of the galaxies of
the universe should be in their cool periods''. They should account for the
low temperature blackbody radiation observed in the CMB.

Since the energy recovery period of a galaxy greatly depends on the
luminosities of its closest neighbors, then the sudden increase of energy
emitted by one of them should accelerate (trigger) the explosion of the
other ones. Thus the regeneration of clusters can occur in ``chains'' that
are consistent with the clusters of luminous galaxies. The clusters of dark
galaxies are consistent with the apparent ``voids'' in the universe.

\section{SOME RECENT ASTRONOMICAL VERIFICATIONS}

\subsection{Galaxies without Dark Matter}

From above, \textit{the minimum proportion of dark bodies should exist in
the recently formed elliptical galaxies}. This fact has been recently
verified by Romanowsky [4]. From his studies of the kinematics of the outer
parts of three intermediate-luminosity elliptical galaxies, he found that
``the data indicates the presence of little if any dark matter in these
galaxies''. ``This unexpected result conflicts with findings in other galaxy
types and poses a challenge to current galaxy formation theories''.

\subsection{Dark Matter Haloes around Galaxies}

During the lifetime of a new galaxy, the halo of dead stars should increase
until the whole galaxy is dark. Thus the x-ray halo coming from hot gas
trapped between the dark bodies can put into relief the existence of such
bodies.

This is consistent with the x-ray halos found around galaxies [5]. The
rather ``spherical'' halo observed around some non-spherical galaxies puts
into relief that the original galaxy was an elliptical one. Thus the
proportion of dark bodies increases with the ``relative'' age of a galaxy.
This accounts for the increase of the proportion of dark matter and the
increase of the central density for dwarf galaxies of smaller luminous
volumes [6].

\subsection{Renewed Stars in New Galaxies}

In the new scenario, the ``renewed stars'', that come from gas captured by
dead stars of the dark period, should become denser and more powerful than
the stars where they come from. They may show some original contamination
from metals formed in the earlier luminous period of the same galaxy, which
would increase with the time. Thus, paradoxically, the ``renewed stars'',
would correspond with the so called ``young stars'' of the current
literature and the ``new stars'' correspond with the ``old stars''. Thus
``the renewed stars in new galaxies'' do correspond with the so called
``young stars in old galaxies'' [7]. Thus the terms ``old'' and ``young''
turn out to be rather ambiguous and misleading. Here, for this reason, they
have been replaced by ``new'' and ``renewed'', respectively .

\subsection{New Galaxies in an Old (Renewed) Universe}

From above, all of the phases of the cyclical evolution of the galaxies
should be present in the sky, anywhere and in any time. They correspond with
recent deep field observations, called ``\textit{Old Galaxies in the Young
Universe}''.

Schade \textit{et al}, for example, have found massive elliptical galaxies
between 8 and 11 billion years ago [8]. ``\textit{There is no evidence for a
decline in the space density of early-type galaxies with look-back time}''.
According to the conventional (hierarchical) model, elliptical galaxies
would be the \textit{oldest} ones and, therefore, they could not exist at
near the presumed beginning of the universe.

D. Schwartz and S. Virani, after observing the quasar SDSSp J1306 at 12.7
billion light years, they found that ``\textit{its X-ray spectrum is
indistinguishable from that of nearby, older quasars}''. ``\textit{Likewise,
its relative brightness at optical and X-ray wavelengths was similar to that
of the nearby group of quasars}'' [9].

D. Farrah et al found that the QSO SDSS J1030, at 12.8 billion light years,
``\textit{appears indistinguishable in any way from lower red shift QSOs,
indicating that QSOs comparable to those seen locally existed less than 1
Gyr after the (presumed) big bang}'' [10].

Since elliptical galaxies and quasars of high redshifts would be the initial
and the final luminous phases of the galaxies, then similar results are
expected for the intermediate phases.

\subsection{The Last Luminous Regions of Galaxies}

Recent astronomical observations clearly show that massive black-holes and
quasars of high red shift are close to the center of galaxies of much lower
red shifts. Obviously, the high difference of red shift is gravitational one
[11]. This would prove that quasars or high redshifts are just the
boundaries of LBH in the central regions of the last luminous regions of
galaxies.

Near the end of the luminous phase of a galaxy, most of the remaining
interstellar plasma of the host dark galaxy should become cool enough to
absorb the high energy radiation emitted by the small quasar. This is
consistent with ``\textit{the paucity of ultraviolet light and higher
frequencies in quasars of red shifts higher than 6.3}'' [12].

After that, for a short period, \textit{the new dark galaxy} could be
detected from the radiation emitted by residual gas, before it is completely
out gassed by its cool bodies. This is consistent with \textit{the two
isolated H I clouds recently found in Virgo} [13].

\subsection{G Time Dilation of Supernovas}

From the NEP, the proportional differences of the relative frequency of the
atoms and of any natural phenomenon that may occur in a NL galaxy, due
either to differences of velocity or due to differences of GP, compared to
those of local galaxies, should be the same. This has been verified by
Goldhaber et al from the identical proportion of the GTD observed in the
light curves of supernovas and of the emission frequency of their atoms [13].

\section{CONCLUSIONS}

A direct test of the NEP comes out from the general correspondence of the
EEP with the results of the genuine GTD experiments done with standard
clocks. This one puts into relief that the original source of the current
problems and errors in physics and cosmology is the CH.

The success of GR comes from the fact that the errors of the GFEH,
introduced by Einstein, are of the same magnitude and opposite sign than the
errors of the CH. Thus, \textit{in the classical tests of GR, the two kinds
of errors are just compensated}, which makes believe in that everything is
OK, which is not true. Because this happens only when the two hypotheses are
used. \textit{The errors stand out when only one of such hypothesis is used}.

The elimination of the two hypotheses bring out clear simplifications,
because the G field has no energy and, therefore, there is not energy
exchange between the G field a body traveling freely in a G field and during
universe expansion. During its free trip, its relative mass, with respect to
any inertial observer, remains constant. The real changes do occur when the
body stops in different GP thus releasing different energies. The
proportional differences of the relative rest masses are just equal to the
corresponding differences of GP.

Then the true reasons for the EEP and for the constant speed of light become
clearer because the real changes occurring after changes of GP or velocity
cannot be detected by observers that have had the same changes of GP or of
velocity. This is because all of the well-defined part of a systems have the
same nature and, therefore, they changed in identical proportion. Then every
local ratio remains unchanged, which makes believe in that nothing has
happened, which is not strictly true.

The new astrophysical scenario fixed by the NEP is also a good test for the
new theory because it has radical differences compared with the standard
scenario, mainly in that the universe expansion cannot produce measurable
changes with the time. Thus the average universe must look like it was
static, i.e., its age may be infinite.

The new kind of ``\textit{linear}'' black hole, without singularity, plays
an important role for the global conservations of the mass-energy and of the
entropy in this universe. It concentrates the dispersed radiation into a
macronucleus that, after a long period, can explode thus generating ``new
primeval gas'' that can transform a dead galaxy into a luminous one and so
on, i.e., it makes possible that matter can evolve, indefinitely, in rather
closed cycles.

Then the galaxies should have been evolving, from long time ago, in rather
closed cycles with luminous and dark periods. Only in this way,
statistically, the different phases of the evolution cycles of the galaxies
can be present in the universe, in the proportion given by their
corresponding periods. The average relative properties of the bodies in the
universe, like their mass-energies, densities and entropies would remain
constants with the time.

Recent astronomical observations clearly verify the existence of all of the
different phases of the galaxy cycles predicted from the NEP, within a wide
region of space-time, up to near the presumed birth of the universe. They
are in clear contradiction with the current hypotheses used for interpreting
the astronomical observations.

Regardless of its high simplicity, the new theory based on properties of
light unifies concepts in several branches of physics, mainly in
gravitation, nuclear physics, quantum mechanics and astrophysics. Thus it
should be considered as a serious alternative for understanding the physical
phenomena occurring in the universe.

\bigskip 

\begin{thebibliography}{99}
\bibitem{1}  R. A. Vera, ``What does the work in a G field?'', Proceedings
of the Einstein's Centennial Symposium on Fundamental Physics, Eds. S. M.
Moore (Universidad de Los Andes, Bogot\'{a}) and G. Violini (Universit\'{a}
di Roma), pp. 597-625, 1981.

\bibitem{2}  R. A. Vera, ``A Dilemma in the Physics of G Fields'',
International Journal of Th. Physics, 20, pp. 19-50, 1981.

\bibitem{3}  R. A. Vera, ``The New Universe Fixed by the EEP and Properties
of Light'', Eds. de la Universidad de Concepcion. Chile \TEXTsymbol{<}%
rvera@udec.cl\TEXTsymbol{>}, 1997

\bibitem{4}  A. J. Romanowsky \textit{et al}, ``A Dearth of Dark Matter in
Ordinary Elliptical Galaxies'', Science, 301 Issue 5640, pp. 1696-1698, 2003.

\bibitem{5}  D. Buote \textit{et al}., ``Chandra Evidence for a Flattened,
Triaxial Dark Matter Halo in the Elliptical Galaxy NGC 720'', Astrophys. J.
577, 183

\bibitem{6}  J. Kormendy, K. C. Freeman, ``Scaling Laws for D. Matter Halos
in Late Type and Dwarf Spheroid Galaxies'', Bull. A. A. S., 30, 1281, 1998

\bibitem{7}  A. Cimatti \textit{et al}, ``Old galaxies in the young
Universe''. Nature 430, 184 - 187 (08 July 2004)

\bibitem{8}  David Schade, ``HST Imaging of the CFRS and LDSS Red shift
Surveys-III. Field elliptical galaxies at 0.2 \TEXTsymbol{<} z \TEXTsymbol{<}
1.0'', ArXiv: astro-ph/9906171.

\bibitem{9}  D. A. Schwartz and S. N. Virani, ``Chandra Measurement of the
X-Ray Spectrum of a Quasar at z = 5.99'', ApJL, 615, L21, 2004. [10] D.
Farrah et al., ``The X-Ray Spectrum of the z = 6.30 QSO SDSS J1030+0524'',
ApJL, 611, L13 G, 2004.

\bibitem{10}  D. Farrah \textit{et al.}, ``The X-Ray Spectrum of the z =
6.30 QSO SDSS J1030+0524'' ApJL, 611, L13 G, 2004.

\bibitem{11}  P. Galianni \textit{et al}, ``The discovery of a high red
shift X-ray emitting QSO very close to the nucleus of NGC 7319 '', ArXiv:
astro-ph/0409215

\bibitem{12}  Becker et al, ArXiv: astro-ph/0108097

\bibitem{13}  R. Minchinar \textit{et al}, ``A Dark Hydrogen Cloud in the
Virgo Cluster''. ArXiv: astro-ph/0502312

\bibitem{14}  Goldhaber \textit{et al}, ``Observation of Cosmological Time
Dilation using Type Ia Supernovae as Clocks'', arXiv: astro- ph/ 9602124.
\end{thebibliography}

\end{document}